\documentclass[a4paper]{jpconf}
\usepackage{graphicx}
\usepackage{epsfig}
\usepackage{amssymb}
\usepackage{amsmath, amscd}
\newcommand{\R}{\mathbb{R}}

\newcommand{\C}{\mathbb{C}}

\usepackage{lipsum}

\newenvironment{itquote}
  {\begin{quote}\itshape}
  {\end{quote}\ignorespacesafterend}

\begin{document}
\title{Random matrices: Application to quantum paradoxes}

\author{Alexey A. Kryukov}

\address{Department of Mathematics \& Natural Sciences, University of Wisconsin-Milwaukee, USA}

\ead{kryukov@uwm.edu}

\begin{abstract}
Recently, a geometric embedding of the classical space and classical phase space of an $n$-particle system into the space of states of the system was constructed and shown to be physically meaningful \cite{KryukovX}. Namely, the Newtonian dynamics of the particles was recovered from the Schr{\"o}dinger dynamics by constraining the state of the system to the classical phase space submanifold of the space of states. A series of theorems related to the embedding and the Schr{\"o}dinger evolution with a random Hamiltonian was proven and shown to be applicable to the process of measurement in classical and quantum mechanics. Here, these results are applied to have a fresh look at the main quantum-mechanical thought experiments and paradoxes and to provide a new insight into the process of collapse and the motion of macroscopic bodies in quantum mechanics. 
\end{abstract}

\section{Introduction}

Here, we apply the theorems proved in \cite{KryukovX} to the analysis of the main quantum-mechanical thought experiments. The following notation, definitions and results will be used. The symbols $M^{\sigma}_{3}$ and $M^{\sigma}_{3,3}$ will denote submanifolds of the projective space of states $CP^{L_{2}}$ of a single particle that consist of all equivalence classes of functions
\begin{equation}
\label{g}
g_{{\bf a}, \sigma}({\bf x})=\left(\frac{1}{2\pi\sigma^{2}}\right)^{3/4}e^{-\tfrac{({\bf x}-{\bf a})^{2}}{4\sigma^{2}}}
\end{equation} 
and
\begin{equation}
\label{phi}
\varphi_{{\bf a}, {\bf p}, \sigma}({\bf x})=\left(\frac{1}{2\pi\sigma^{2}}\right)^{3/4}e^{-\tfrac{({\bf x}-{\bf a})^{2}}{4\sigma^{2}}}e^{i{\bf p}{\bf x}/\hbar}
\end{equation} 
respectively.
Here, ${\bf x}, {\bf a}$ and ${\bf p}$ are vectors in $\R^3$ and the variance $\sigma^2$ is assumed to be sufficiently small.
The space of states $CP^{L_{2}}$ is furnished with the Fubini-Study metric.
The submanifolds $M^{\sigma}_{3}$ and $M^{\sigma}_{3,3}$ possess the induced Riemannian metric. The map $\omega: \R^3 \longrightarrow M^{\sigma}_{3}$ defined by $\omega({\bf a})=g_{{\bf a}, \sigma}$ identifies the Euclidean space $\R^3$ with the Riemannian manifold $M^{\sigma}_{3}$. Likewise, the map $\Omega: \R^6 \longrightarrow M^{\sigma}_{3,3}$, $\Omega({\bf a}, {\bf p})=\varphi_{{\bf a}, {\bf p}, \sigma}$ identifies the classical Euclidean phase space $\R^6$ with the Riemannian manifold  $M^{\sigma}_{3,3}$.

The functions $g_{{\bf a}, \sigma}$ and $\varphi_{{\bf a}, {\bf p}, \sigma}$ represent well-localized single-particle wave packets with group velocities zero and ${\bf p}$ respectively. The theorem ${\bf (A)}$ in \cite{KryukovX} asserts that the Schr{\"o}dinger dynamics with the state function constrained to the classical phase space manifold $M^{\sigma}_{3,3}$ is equivalent to the Newtonian one. In the case of an $n$-particle system, the same is true provided the state of the system is constrained to the product $M^{\sigma}_{3n,3n}=M^{\sigma}_{3,3}\otimes ... \otimes M^{\sigma}_{3,3}$ of $n$-copies of the manifold $M^{\sigma}_{3,3}$.

A repeated measurement of the position of a macroscopic particle can be modeled by a random walk on $\R^3$ with normally distributed steps. By the end of observation period, the position random variable is normally distributed, in agreement with the observations and the central limit theorem. The random walk without drift on $\R^3$  admits a unique extension to a random walk on the space of states whose steps satisfy the Schr{\"o}dinger equation with the Hamiltonian given by a random matrix from the Gaussian unitary ensemble. The conjecture ${\bf (RM)}$, based on the generally accepted BGS-conjecture \cite{BGS}, is the assumption that such a random walk in the space of states models the behavior of the corresponding quantum-mechanical system under a measurement. 
Theorem ${\bf (E)}$ in \cite{KryukovX} states that the probability to find an initial state at a certain end-point as a result of the described random walk is given by the Born rule. It follows that the walk can indeed be used to model the process of measurement of the position in classical and quantum physics. 


When the position of a macroscopic particle is measured, the fluctuation from the mean in the observed values in different trials can be associated with the particle's interaction with the surroundings. The atoms and molecules of the medium in thermal motion kick the measured particle and cause its displacement. The random walk that models the process of observation approximates in this view the resulting Brownian motion of the particle in the medium. Because a displacement of the measured particle results in the like-displacement of its state, the random walk in the space of states in the conjecture ${\bf (RM)}$ must be the unique extension of the random walk on $\R^3$ that models the classical measurement. This tying of the random walk on $\R^3$ and in the space of states is consistent with the theorem ${\bf (A)}$ that identifies the Schr{\"o}dinger and Newtonian evolutions of the particle whose state is constrained to $M^{\sigma}_{3}=\R^3$.

The above observations result in the theorem ${\bf (F)}$ that claims that the state of a particle is constrained to $M^{\sigma}_{3}$ exactly when the particle is sufficiently large in size, so that the corresponding Brownian motion of the particle is negligible. 
By applying this theorem to a system consisting of a microscopic particle and a macroscopic measuring device, one obtains theorem ${\bf (G)}$. It claims that the state of such a system will maintain its product form (i.e., with no entanglement between the particle and the device) with the device behaving classically during the measurement.

In what follows, we begin with a discussion of the experiments involving a single microscopic particle, followed by experiments with two entangled microscopic particles and then consider experiments involving microscopic and macroscopic systems. In doing so, we reformulate the experiments in geometric terms and apply the above-mentioned theorems. At the end of the paper, we demonstrate how the Born rule is a generic consequence of the Schr{\"o}dinger evolution with random Hamiltonian and review the motion of macroscopic bodies in quantum mechanics.

\section{The double-slit and delayed-choice experiments}
\label{double}


Feynman once famously said that the double-slit experiment ``...is a phenomenon which is impossible, absolutely impossible, to explain in any classical way, and which has in it the heart of quantum mechanics. In reality, it contains the only mystery'' \cite{Feynman}. The paradox of the double-slit experiment and the Wheeler's delayed-choice versions of it is in the apparent double nature of the photon or electron in the experiment. Somehow, the particle behaves as a wave or as a particle, depending on a change in the configuration of the experiment, even if the change happened after the particle in the experiment already went ``through" the slits or the beam-splitter. So the nature of the particle is somehow decided at the last moment by the device used to observe it. In Wheeler's words, the ``past has no existence except as recorded in the present" and the universe does not ``exist, out there independent of all acts of observation''.

Let us first interpret the property of being a particle or a wave in terms of the geometry of the manifold $M^{\sigma}_{3}$ and the ambient space of states. We see the electron as a particle when its state is a point {\it on} or {\it near} $M^{\sigma}_{3}$. We see the electron as a wave when its state is {\it away} from $M^{\sigma}_{3}$.
In fact, to behave like a particle is to have a state that is well localized, that is, close to $M^{\sigma}_{3}$ in the Fubini-Study metric on the space of states. To behave like a wave is to have a state function that is sufficiently wide-spread in the coordinate representation, which means that the state is away from $M^{\sigma}_{3}$. 

For instance, consider a Gaussian state $\varphi$ with $|\varphi|^2$ of variance $d^2$, centered at a point ${\bf a}$ in $\R^3$ and the corresponding state $g_{{\bf a}, \sigma}$ closest to it in $M^{\sigma}_{3}$. We have for their inner product in $L_{2}(\R^3)$
\begin{equation}
\label{gg}
(\varphi, g_{{\bf a}, \sigma})=\left(\frac{2\sigma d}{\sigma^2+d^2}\right)^{\frac{3}{2}}.
\end{equation}
If $d=\sigma$, the right hand side is equal to $1$, which means that the Fubini-Study distance between the states is $0$: the states are the same. As $d$ increases, so does the distance between the states. If $d \gg \sigma$, the right side of (\ref{gg}) approaches $0$. That is, distance between the states approaches $\pi/2$, which is the largest possible distance on $CP^{L_{2}}$. This describes the general situation: the larger the spread of $|\varphi|^2$, the larger is the distance between $\varphi$ and $M^{\sigma}_{3}$.

In the double-slit experiment, the state of the particle at the source is well-localized, so we can assume that it approaches the screen with the slits as a particle. Upon interaction with the screen, the state of the particle is a superposition of a pair of localized wave packets, each corresponding to the particle passing through one of the slits. This means that the variance of $|\varphi|^2$ has increased, so the Fubini-Study distance from the superposition to $M^{\sigma}_{3}$ has increased as well. The state is moving away from the classical space $M^{\sigma}_{3}=\R^3$ into the space of states. The particle is becoming a wave. We can verify this property by inserting a photographic plate and observing the interference pattern on it.

However, when the electron hits the photographic plate, it shows up as a particle. So how did the wave become a particle again?  Note that a classical particle  hitting a plate experiences a random motion. This motion is due to the particle's interaction with atoms and molecules of the plate and is similar to the motion of a bullet in a sandbox. As already discussed, such a motion can be modeled by a random walk that yields Gaussian distribution of the end-position of the particle. According to ${\bf (RM)}$, the state of a microscopic particle exposed to similar conditions shall experience a random walk in the space of states.  By theorem ${\bf (E)}$, the probability that the state will reach a particular point of the plate during the walk is given by the Born rule, which is consistent with observation. So, under interaction with the plate, the state $\varphi$ of the electron wanders around the space of states before reaching a point on the plate. Reaching the plate implies reaching the manifold $M^{\sigma}_{3}$, which means that the state of the electron is now well localized. The electron in the experiment demonstrates now properties of a particle. 

To be more precise, the plate is a collection of a large number $n$ of particles each having a reasonably well defined position. 
%
That is, the plate is given by a point $\psi_P$ on the submanifold $M^{\sigma}_{3n}=M^{\sigma}_{3} \otimes ... \otimes M^{\sigma}_{3}$ in the space of states $CP^{L_{2,n}}$. Here $L_{2,n}$ is the tensor product of Hilbert spaces $L_{2}(\R^3)$ for all particles of the plate. The isomorphism $\omega_{n}: \R^3 \times...\times \R^3 \longrightarrow M^{\sigma}_{3n}$, $\omega_{n}({\bf a}_{1}, ... , {\bf a}_{n})=g_{{\bf a}_{1}, \sigma}\otimes ... \otimes  g_{{\bf a}_{n}, \sigma}$ allows us to view the states in $M^{\sigma}_{3n}$ as points in the classical configuration space $\R^{3n}$  or positions of $n$ particles in the single classical space $\R^{3}$. That is how our usual view of the plate becomes possible and how the state $\psi_P$ gets identified with a set of material points that represent the particles of the plate in $\R^3$. Because the plate is macroscopic, theorem ${\bf (G)}$ applies, telling us that the state of the particle-plate system is a product $\varphi \otimes \psi_P$. Because of its product form, the state of the system is close to $g_{{\bf b}, \sigma} \otimes \psi_P$ in the space of states of the system exactly when the state $\varphi$ is close to the state $g_{{\bf b}, \sigma}$ in the space $CP^{L_{2}}$ of the particle.

What happens if the choice of configuration of the experiment is delayed and occurs while the particle is already on its way? Suppose for example that we decided to detect which slit the particle went through by inserting a detection screen after the particle went through the slits, but before it encounters the photographic plate. In this case, the first part of the particle's path remains the same. After the screen with the slits, the state $\varphi$ of the particle is a superposition of two Gaussian-like states, which means, as discussed, that $\varphi$ moved away from $M^{\sigma}_{3}$. 
The interaction of the particle with the detection screen happens in the same way as its interaction with the plate. Since the corresponding classical system experiences a random walk, the conjecture ${\bf (RM)}$ applies, predicting a random walk of $\varphi$ in the space of states and the resulting Born rule for the probability of the state to be found by one of the slits. If the state is found by one of the slits, then this means that it is now in a well-localized form and therefore the electron behaves as a particle. The interference effect on the photographic plate disappears.

\section{``Faster-than-light"}
\label{faster}

Consider a microscopic particle confined to a box (infinite potential well). In quantum mechanics, as soon as the box is open, the wave function of the particle instantly spreads over the entire space \cite{Hegerfeldt}. 
As is well known, the probabilistic nature of the wave function with regard to observations precludes superluminal signaling based on this and the like phenomena. Still, it is not clear what should we make of them.

Suppose the closed box is sufficiently small and located at a point ${\bf a}$ in $\R^3$. Then the state $\varphi$ of the particle in the box is close in the Fubini-Study metric to the point $g_{{\bf a},\sigma}$ in $M^{\sigma}_{3}$. As soon as the box is open, the distance between $\varphi$ and $g_{{\bf a},\sigma}$ {\it increases} and quickly approaches $\pi/2$, which is the largest possible distance in the space of states $CP^{L_2}$ of the particle. At the same time, the distance between $\varphi$ and the points $g_{{\bf b},\sigma}$ with ${\bf b}$ outside the box, {\it decreases} from the initial value of $\pi/2$ to a smaller value. To visualize it, recall that the space $M^{\sigma}_3$ spirals though dimensions of the space of states and forms an overcomplete basis in this space. The distance between two points $g_{{\bf a},\sigma}$  and $g_{{\bf b},\sigma}$ in $M^{\sigma}_{3}$ with $||{\bf a}-{\bf b}||_{\R^3}\gg \sigma$ is close to $\pi/2$. When the particle ``escapes" the box, the point $\varphi$, representing the particle's state, moves away from $M^{\sigma}_{3}$ approaching the points  $g_{{\bf b},\sigma}$ in $M^{\sigma}_{3}$, with ${\bf b}$ outside the box. That is because the geodesics connecting $\varphi$ and $g_{{\bf b},\sigma}$ is shorter than the geodesics connecting $g_{{\bf a},\sigma}$ and $g_{{\bf b},\sigma}$. The latter distance is  still much shorter than the Euclidean distance $||{\bf a}-{\bf b}||_{\R^3}$, measured along the ``spiraling" submanifold $M^{\sigma}_{3}$. 

Suppose now that we decided to measure the position of the particle that escaped from the box. According to ${\bf (RM)}$, such a measurement results in a random walk of the particle's state and the Born rule for the probability of a particular observed position. Because the distance from $\varphi$ to  $g_{{\bf b},\sigma}$ has decreased, the probability to find a particle at ${\bf b}$ is now positive for all ${\bf b}$, even those far away from ${\bf a}$. So, we may be able to find the particle at a sufficiently distant point ${\bf b}$ for which the motion from ${\bf a}$ to ${\bf b}$ in $\R^3$ during the observation would need to be superluminal. This seems to be paradoxical because we imagine the particle moving, in fact, through $\R^3$ that is, through $M^{\sigma}_{3}$. However, the state of the particle does not move through $M^{\sigma}_{3}$ but rather through the space of states. So, there is no superluminal motion in $\R^3$, rather just a motion in the space of states, not confined to $M^{\sigma}_{3}$. Confusing these two motions is what leads to the paradox. As already mentioned, the probabilistic nature of the motion of state under measurement precludes using the effect to transmit a signal from point ${\bf a}$ to point ${\bf b}$. 

 Let us also point out that the ``size" (radius) of the sphere of unit-normalized states in the space of states has nothing to do with the ``size" of the classical space $\R^3$. The sphere can be miniscule and yet the entire classical space $\R^3$ can be isometrically embedded into it. That is because under the embedding the infinite size of $\R^3$ is realized through infinite dimensions of the sphere, and not through its radius. In fact, if ${\bf a} \neq {\bf b}$, then $g_{{\bf a},\sigma}$ and $g_{{\bf b}, \sigma}$ are linearly independent. It is therefore clear that the speed of motion of state in the space of states does not directly correspond to the speed of a process in the classical space. The physical motion of state may be as slow as desired, while the speed of an apparent ``communication" between regions in $\R^3$ may be superluminal.


\section{EPR-paradox}

The Einstein-Podolsky-Rosen thought experiment \cite{EPR} deals with two paradoxical properties of an entangled pair of particles. First, the position or momentum property of one of the particles in the EPR-pair determines reality of one, but not the other corresponding property of the second particle. This is paradoxical: How could the second particle change its nature because of a measurement, particularly when the measurement is done on a different particle? Second, the ``communication" needed to convey the result of measurement of the first particle to the second one needs to be nearly instanteneous. For particles sufficiently distant from each other, it must be faster than light, seemingly violating causality. 

Let us again interpret the situation geometrically. The particles have a definite position when the state of the pair belongs to the submanifold $M^{\sigma}_{3} \otimes M^{\sigma}_{3}$ of the space of states of the pair. The particles have a definite momentum when the state of the pair belongs to the submanifold $\widetilde{M}^{\sigma}_{3} \otimes \widetilde{M}^{\sigma}_{3}$, where $\widetilde{M}^{\sigma}_{3}$ is the Fourier image of $M^{\sigma}_{3}$. Since for sufficiently small $\sigma$, these two manifolds do not intersect, the position and momentum of the particles cannot exist at the same time. Instead of a simultaneous ``reality" of the position and momentum of the particles, physical properties of the pair are always determined by the state of the pair.  Only when the state belongs to appropriate submanifolds of the space of states, the Newtonian physical quantities appear and make sense.

Let us apply this to measurements on the EPR-pair of particles. According to ${\bf (RM)}$, a measurement of the position of one of the particles in the EPR-pair results in a random motion of the initial state of the pair and the Born rule for the probability of transition to a certain end-state. If the state reaches the manifold $M^{\sigma}_{3} \otimes M^{\sigma}_{3}$, the positions of the first and second particles will be real, while the momentum will not. If the momentum of one of the particles is measured, the state will be found on the manifold $\widetilde{M}^{\sigma}_{3} \otimes \widetilde{M}^{\sigma}_{3}$, making the momentum of both particles real, while their position will not exist. Between the measurements, the state of the pair is on neither of these manifolds and so neither position, nor momentum of the particles is real.  In all of these cases, the state of the pair identifies physical reality of the pair.

What about the seemingly instant communication between the particles? This question can be answered in a way similar to the case of a single particle (see Section \ref{faster}).  Namely, when the pair has a definite position, its state is $\varphi={\widetilde \delta}^3_{\bf a}\otimes {\widetilde \delta}^3_{{\bf a}+{\bf \Delta}}$, where ${\bf \Delta}$ is a vector in $\R^3$. This state is a superposition of states of a definite momentum, i.e., states in $\widetilde{M}^{\sigma}_{3} \otimes \widetilde{M}^{\sigma}_{3}$. To measure the momentum of one (and therefore both) of the particles is to bring the state $\varphi$ to the manifold  $\widetilde{M}^{\sigma}_{3} \otimes \widetilde{M}^{\sigma}_{3}$. Such a motion is not happening in the classical space $M^{\sigma}_{3} \otimes M^{\sigma}_{3}$ of the pair, but in the space of states. Therefore, it has nothing to do with a signal in the classical space. It is then meaningless to talk about superluminal communication between the particles. There is none. The reality is given by the state of the pair and its evolution in the space of states, and not by two particles in the classical space. Only when the state belongs to the submanifolds $M^{\sigma}_{3} \otimes M^{\sigma}_{3}$ or $\widetilde{M}^{\sigma}_{3} \otimes \widetilde{M}^{\sigma}_{3}$ the notion of the particles emerges and the classical language becomes applicable to the pair.

\section{Schr{\"o}dinger's cat experiment}

The Schr{\"o}dinger's cat thought experiment appeared in the discussion of Schr{\"o}dinger and Einstein about the nature of quantum superpositions.  Schr{\"o}dinger proposed the experiment to demonstrate that the issues related to quantum superpositions are not limited to microscopic systems. For Einstein, the experiment was a perfect illustration of the need to extend the notion of macroscopic reality into quantum domain, as he and the coauthors argued earlier in \cite{EPR}. 
For both scientists, the outcome of the experiment was absurd, leading to a paradox that needed to be addressed. Most of the later authors seemed to accept the outcome as real and attempted to interpret it one way or another.

Let us apply the theorem ${\bf (G)}$ in  \cite{KryukovX} to the experiment. Suppose that at the beginning of the experiment, the state of the cat belongs to the manifold $M^{\sigma}_{3}$. Suppose also that the atom-cat system is initially in a product state and that the conjecture  ${\bf (RM)}$ is valid.  The theorem claims that in this case, given the macroscopic nature of the cat, the state of the atom-cat system maintains its product form throughout the experiment. The theorem also claims that the cat will behave in accord with the laws of classical mechanics, while the atom will behave quantum-mechanically. 
So the assumptions of theorem  ${\bf (G)}$ are sufficient to resolve the paradox.

\section{Wigner's friend}

Wigner has demonstrated that the rules of quantum mechanics may lead to contradictory accounts of observations performed by two observers.  In his thought experiment \cite{Wigner2}, an observer (Wigner's friend) makes a measurement on a system and another observer (Wigner himself) observes the laboratory, where his friend performs the measurement. The contradiction is seen in the observers getting two different states after the friend has performed the measurement, but before the result is communicated to Wigner. This, as well as more recent thought experiments based on a similar scenario, raise the question of when the collapse of the state actually occur, what does the consciousness of the observers have to do with the collapse, and, ultimately, is there such a thing as an objective reality that all observers can agree on.

The paradox of the Wigner's friend thought experiment is tied to the Schr{\"o}dinger cat paradox and can be addressed in a similar way. Namely, assuming the conditions of the theorem ${\bf (G)}$ are met, Wigner, the friend and the measured system can only exist in a product state, making the paradox disappear. 
Note that the observer and the lab states are constrained to the manifold $M^{\sigma}_{3}$, identified with the classical space we live in.
We therefore have a very special frame of reference to describe events happening on a much larger manifold of states. There is a whole new world out there, available only to microscopic systems, where a different kind of reality exists.  This new, microscopic reality complements the reality we know, and is applicable to systems capable of escaping the classical space $M^{\sigma}_{3}$.

\section{The Born rule and random matrices}

The random walk description of the process of collapse is an essential part of the analysis of the thought experiments in the paper. It is therefore important to provide a better understanding of the collapse, beyond the abstract statements of theorems in \cite{KryukovX}. 
%
The random walk under consideration is defined by the following two conditions: 
(1) The steps of the random walk without drift satisfy the Schr{\"o}dinger equation with the Hamiltonian represented by a random matrix from the Gaussian unitary ensemble (GUE). (2) Matrices representing the Hamiltonian at two different times are independent and belong to the same ensemble. 

Here is what can be derived from this definition.
First of all, the random walk under consideration happens on the space of states $CP^H$. 
At any point $\{\varphi\}$ in the space of states, the step $d\varphi=-\frac{i}{\hbar}{\widehat h}\varphi dt$ is a normal random vector with spherical distribution in the tangent space $T_{\{\varphi\}}CP^H$. The steps are mutually independent random vectors. The probability of reaching a point in $CP^H$ starting from a given initial point depends only on the Fubini-Study distance between the points. Under the condition that the walk takes place on the submanifold $M^{\sigma}_{3}$, the walk is the usual random walk with Gaussian steps on $\R^3$. In particular, the position random vector is normal. Without the constraint to $M^{\sigma}_{3}$, the probability of reaching a point in $CP^H$ is given by the Born rule.

Informally speaking, the state of the measured system is kicked around the space of states, with no privileged direction. Somehow, there must be a non-vanishing probability to find the state at or near an eigenstate of the measured observable. How is it possible? First of all, the dimension of the space of states in any measuring experiment is finite. For example, suppose we want to measure the position of a particle on the photographic plate in the double-slit experiment. At the very best, we can identify the molecule or atom ionized by the particle in the experiment. Since there are finitely many atoms on the surface of the plate, the space of states for the experiment is finite-dimensional. Therefore, the volume form and the probability density function are well defined on the space. 
However, the volume of a cubical cell in $\R^n$ is the $n$-th power of its side. So, the probability of finding the particle in a small cell decreases exponentially with the dimension of space. So the question remains. 

The adequate answer lies in the clever design of our measuring devices. 
Consider, once again, the double slit experiment with photographic plate. 
Although the form of the initial wave packet of the particles at the source needed to observe interference pattern is not particularly important, the packet must propagate towards the screen with the slits and the photographic plate. That means that the path of the particle's state in the space of states must approach the screen and the plate in the Fubini-Study metric. 
In other words, the device must create a ``drift" of state towards the manifold $M^{\sigma}_{3}$ and towards the screen and the plate in it. It is important to remind the reader that the space of states has very high dimensionality while $M^{\sigma}_{3}$ had dimension $3$, so that this condition is by no means trivial. Even for a spherical screen that surrounds the source, there are infinitely many linearly independent directions in the space of states, such that the lines from the source in these directions never get close to the screen. So how does the Schr{\"o}dinger dynamics of the particle interacting with the device explains the drift?

It was argued by Born \cite{Born}, Heisenberg \cite{Heis} and Darwin \cite{Darwin}, and soon after proved by Mott in his paper on $\alpha$-ray tracks \cite{Mott} that under proper conditions, the Schr{\"o}dinger dynamics is capable of explaining straight tracks of particles in a medium. Mott demonstrated that given the right conditions, excitation of several atoms of the medium resulting from a passage of the scattered particle is much more likely when the atoms and the particle's source lie on the same straight line. The missing part in deriving the classical path of the particle in this scenario was an explanation of how a particular straight track out of the many possible ones was selected. Expanding on Mott's approach, the phenomenon of decoherence was more recently used to explain the transition from quantum to classical in a variety of phenomena. Here too, the theory stopped short of explaining how a particular classical reality out of all possible ones is selected.

On another hand, the random walk without drift in the version  ${\bf (RM})$ of the BGS-conjecture explains how a particular outcome of a measurement may appear and provides the probability of the outcome. The probability is {\it conditional}, given that the outcome is an eigenstate of the measured observable. Complementary to this, Mott's result indicates how the device configuration may be responsible for fulfilling this condition. Namely,
interaction of the particle with the medium analyzed in \cite{Mott} and entirely based on the Schr{\"o}dinger dynamics may explain the drift towards the classical space manifold $M^{\sigma}_{3}$ in the double-slit experiment. As the particle goes through the emulsion of the photographic plate and excites electrons of a chain of atoms, it loses energy and eventually gets absorbed by a particular atom. At this point, the state of the particle is localized and position of the particle is determined. 
%
By decreasing the ``spread" of state, the device eliminates the end-states that are away from $M^{\sigma}_{3}$, for example, the eigenstates of the momentum operator. So, the design of the measuring device and the usual Schr{\"o}dinger dynamics are responsible for ``killing" alternative outcomes of the experiment.

Mathematically, the drift for a random walk is a vector that is added to steps of the walk. 
Note that for the case under consideration, the walk in the directions orthogonal to the drift vector is unchanged. For instance, let the path through the state (\ref{g}) be obtained by changing the variance $\sigma^2$. It can be shown to be orthogonal to $M^{\sigma}_{3}$ (see \cite{KryukovX}).  Consider the random walk specified in ${\bf (RM)}$ with the drift vector tangent to such a path at the photographic plate. Then the distribution of steps of the random walk with drift along the surface of the plate will not change. In particular, the Born rule for the probability of finding the particle at a particular point of the plate will remain valid. 

\section{Schr{\"o}dinger dynamics of macroscopic bodies}

Let us review how in the developed theory the Newtonian motion of macroscopic particles is derived from the Schr{\"o}dinger dynamics of the particles. Under interaction with the surroundings, the state of a macroscopic particle experiences a random walk on the space of states. The walk conditioned to stay on the classical space submanifold $M^{\sigma}_{3}$ models the usual Brownian motion of the particle. For sufficiently large particles, the Brownian motion and the random walk in the space of states trivialize. In the absence of an external potential, the state of such a particle is at rest in the lab system in $M^{\sigma}_{3}$. In particular, the state does not spread. By considering the state in the coordinate system moving relative to the lab, i.e., by applying the operator of the Galileo transformation to the state, we conclude that the transformed state will be contained in the classical phase space submanifold $M^{\sigma}_{3,3}$. Observed from the coordinate system that is translationally accelerated with respect to the lab system, the state will also belong to $M^{\sigma}_{3,3}$ (see \cite{Mashhoon} and the references therein). The Schr{\"o}dinger equation in the accelerated system contains an inertial potential term \cite{Mashhoon}. In the linear approximation appropriate for the states in $M^{\sigma}_{3,3}$ with a small $\sigma$, the inertial potential is equivalent to an external potential. It follows that under an external potential, the state of the particle will be confined to $M^{\sigma}_{3,3}$. The theorem ${\bf (A)}$ then guarantees that the motion will be the usual Newtonian motion of the particle in the potential.





\end{document}